\documentclass[preprint2]{aastex}
\usepackage{natbib}

\shorttitle{Frequency independent quenching of pulsed emission}
\shortauthors{Gajjar et al.}

\begin{document}

\title{Frequency independent quenching of pulsed emission}

\author{Vishal Gajjar}
\affil{National Centre for Radio Astrophysics (TIFR), Post Bag 3, 
    Ganeshkhind, Pune - 411007 India; \\
    Xinjiang Astronomical Observatory, Chinese Academy of Science, 
    40-5 South Beijing Road, Urumqi, Xinjiang, 830011, China}
\email{gajjar@ncra.tifr.res.in}

\author{Bhal Chandra Joshi}
\affil{National Centre for Radio Astrophysics (TIFR), Post Bag 3, 
    Ganeshkhind, Pune - 411007 India}
\email{bcj@ncra.tifr.res.in}

\author{Michael Kramer}
 \affil{Max Planck Institute for Radio Astronomy
 Auf dem Hugel 69, P.O. Box 20 24, D-53010 Bonn, Germany; \\ 
 Jodrell Bank Centre for Astrophysics, University of Manchester, 
 Alan-Turing-Building, Manchester M13 9PL, UK}
\email{mkramer@mpifr-bonn.mpg.de}

\author{Ramesh Karuppusamy}
\affil{Max Planck Institute for Radio Astronomy
 Auf dem Hugel 69, P.O. Box 20 24, D-53010 Bonn, Germany}
\email{ramesh@mpifr-bonn.mpg.de}
 
\and

\author{Roy Smits}
\affil{Jodrell Bank Centre for Astrophysics, School of Physics and
  Astronomy, University of Manchester, Manchester M13 9PL, UK; \\ 
Stichting ASTRON, Postbus 2, 7990 AA Dwingeloo, the Netherlands}
\email{smits@astron.nl}

\begin{abstract}
Simultaneous observations at four different frequencies 
\emph{viz.} 313, 607, 1380 and 4850 MHz, 
for three pulsars, PSRs B0031$-$07, B0809+74 and B2319+60, 
are reported in this paper. Identified null and burst pulses 
are highly concurrent across more than decade of frequency. 
Small fraction of non-concurrent pulses ($\leq$3\%) are observed, 
most of which occur at the transition instances. 
We report, with very high significance for the first time, 
full broadband nature of the nulling phenomenon in these three pulsars. 
These results suggest that nulling invokes changes on the global 
magnetospheric scale. 
\end{abstract}


\section{Introduction}
Pulsars show a variety of pulse to pulse variation in 
their pulsed emission. The most dramatic variation 
seen in many pulsars, with emission in a burst of one 
to several pulses interspersed with pulses with  
no detectable radio emission, is called pulse-nulling 
\citep{bac70}. This phenomenon has been reported in more than 100 
pulsars to date \citep{rit76,ran86,big92a,viv95,wmj07,gjk12}. 
The pulsed emission abruptly declines by more than two orders of 
magnitudes during pulsar nulls \citep{la83,vj97,gjk12}, 
which are as yet not well understood. 
Several sensitive studies of nulling pulsars 
have estimated the fraction of pulses with no detectable emission,  
called nulling fraction (NF). These range from 
less than a percent to about 90 percent \citep{rit76,wmj07,gjw13}. 
Most of these studies were carried out at a single 
observing frequency with a typical observation length 
of about an hour. There are very few long simultaneous 
observations of nulling pulsars reported so far. 

In a simultaneous single pulse study 
of two pulsars, PSRs B0329+54 and B1133+16 at 327 and 2695 MHz, 
\cite{bs78} showed highly correlated pulse energy fluctuations, 
indirectly suggesting that nulling is a broadband phenomenon 
within this frequency range. However, only half of nulls 
were reported to occur simultaneously at 325, 610, 1400 
and 4850 MHz for one of these pulsars, PSR B1133+16 \citep{bgk+07}.
\cite{tmh75} reported simultaneous nulls at 275 and 430 MHz for PSRs 
B0031$-$07 and B0809+74. Two separate studies on PSR B0809+74 reported 
contrary broadband nulling behavior. \cite{bkk+81} reported concurrent 
nulling behavior at 102 and 1720 MHz for  
two long nulls while such concurrent behaviour 
was not apparent for three single period nulls, 
probably due to low signal-to-noise ratio (S/N). 
However, in simultaneous observations at 102 and 408 MHz, 
\cite{dls+84} reported highly non-concurrent behavior, 
as only 3 out of 9 nulls were shown to be simultaneous. 

There are two different group of theories  
that attempt to explain the cause of nulling. 
Intrinsic effects such as cessation of primary particles 
on short time-scale \cite{klo+06}, loss of coherence conditions \cite[]{fr82,zqh97b} 
or changes in the current flow \cite[]{tim10} may cause 
a pulsar to null. However, geometric effects such 
as line-of-sight passing between the emitting sub-beams 
producing the so-called {\itshape pseudo-nulls} \cite[]{hr07,hr09, rw08} 
or changes in the direction of the entire emission beam \cite[]{aro83c,gle90,dzg05,zx06}
can also mimic absence of detectable emission. 
As the emission at various radio frequencies is believed to originate at 
distinct heights from the neutron star surface \cite[]{kom70,cor78,mr02a}, 
the spacing between the sub-beam, in the rotating carousal, 
may also differ significantly from lower to higher frequencies 
(However, this needs to be confirmed firmly with more observations). 
Thus, the existence of 
{\itshape pseudo-nulls} \cite[]{hr07} can be tested with simultaneous multiple frequency observations. 
Previously studied pulsars such as PSRs B0809+74 
and B1133+16 represent a model of conal cut pulsar beam \citep{rr03,ran93}, 
where it is difficult to distinguish between geometric and 
intrinsic effects.   
More long, sensitive, and  
simultaneous observations at multiple frequencies 
of a carefully selected sample of pulsars are 
motivated to distinguish between the above mentioned two 
groups of models.

In this paper, we report on long simultaneous 
multi-frequency observations of three pulsars, 
PSRs B0031$-$07, B0809+74 and B2319+60 to 
investigate the broadband nature of pulse 
nulling. These pulsars were chosen as 
(a) they are strong pulsars allowing an 
easy determination of nulls, (b) two of 
these pulsars, PSRs B0031$-$07 and B2319+60, 
show long prominent nulls \citep{htt70,rit76,wf81,viv95}
and have high NF (40\% and 30\%), 
(c) two of these pulsars, PSRs B0031$-$07 
and B0809+74, show prominent drifting and single 
component profile \citep{la83,vj97}
indicating a tangential and peripheral 
line-of-sight traverse of their emission beam, 
and (d) PSR B2319+60 shows a multiple component 
profile with drifting in outer components suggesting  
a more central line of sight traverse of the emission 
beam \cite[]{ran86,wf81,gl98}. Thus, this 
sample allows us to test the effect 
of pulse nulling as a function of observational 
frequency for different parts of pulsar beam 
and discriminate between a geometric or 
an intrinsic origin for pulse nulling. In Section 
\ref{obsanal}, observations and analysis procedures 
are described. The results are presented in 
Section \ref{res}. The conclusions of the 
study are presented in Section \ref{conc} 
along with a discussion on the implications 
of the results. 
   
\section{Observations and Analysis} 
\label{obsanal}
\begin{table*}[h!]
\begin{center}
\caption{Parameters of the observed pulsars and details of observations}
\label{tabobs}
\begin{tabular}{l|c|c|l|c|c}
\tableline\tableline
Pulsar & Period & Dispersion     & Date of       & Number of pulses & Frequencies\\
       &        & Measure        & observations  &              & of observations    \\
\tableline
       &  (s)   &(pc\,cm$^{-3}$) &                &            & (MHz)      \\
\tableline
PSR B0031$-$07 & 0.942951 & 11.38  & 2011 February 5  &  10441 & 313, 607, 1380\\
PSR B0809+74   & 1.292241 &  6.12  & 2011 February 17 &  10003 & 313, 607, 1380, 4850\\
PSR B2319+60   & 2.256488 & 94.59  & 2011 February 6  &  5126 & 313, 607, 1380, 4850 \\
\tableline
\end{tabular}
\end{center}
\end{table*}

PSRs B0031$-$07, B0809+74 and B2319+60 were observed simultaneously 
with the Giant Meterwave Radio Telescope (GMRT), 
the Westerbork Synthesis Radio telescope (WSRT) and 
the Effelsberg Radio Telescope. The details of observations 
are given in Table \ref{tabobs}. The GMRT \citep{sak+91} 
was used in a phased array mode with two sub-arrays 
consisting of about nine 45-m antennas each covering 
313 and 607 MHz band respectively with 33-MHz bandwidth.
The phased array output for each of the two frequencies 
were recorded with 512 channels over the passband using the GMRT 
software baseband receiver with an effective sampling 
time of 1 ms along with a time stamp for the first recorded 
sample, derived from a GPS disciplined Rb frequency 
standard. The variations in the ionospheric and 
instrumental delays across the GMRT sub-arrays have a 
typical time scale of about 2 hours at the observed 
frequencies. Hence, the observations were divided 
into 4 to 5 observing sessions, each of 2 hours, 
interspersed with compensation for the instrumental delay 
drift to maintain phasing of the sub-array. 

\begin{figure}
\begin{center}
\centering
\includegraphics[width=3.3in,height=7.05in,angle=0]{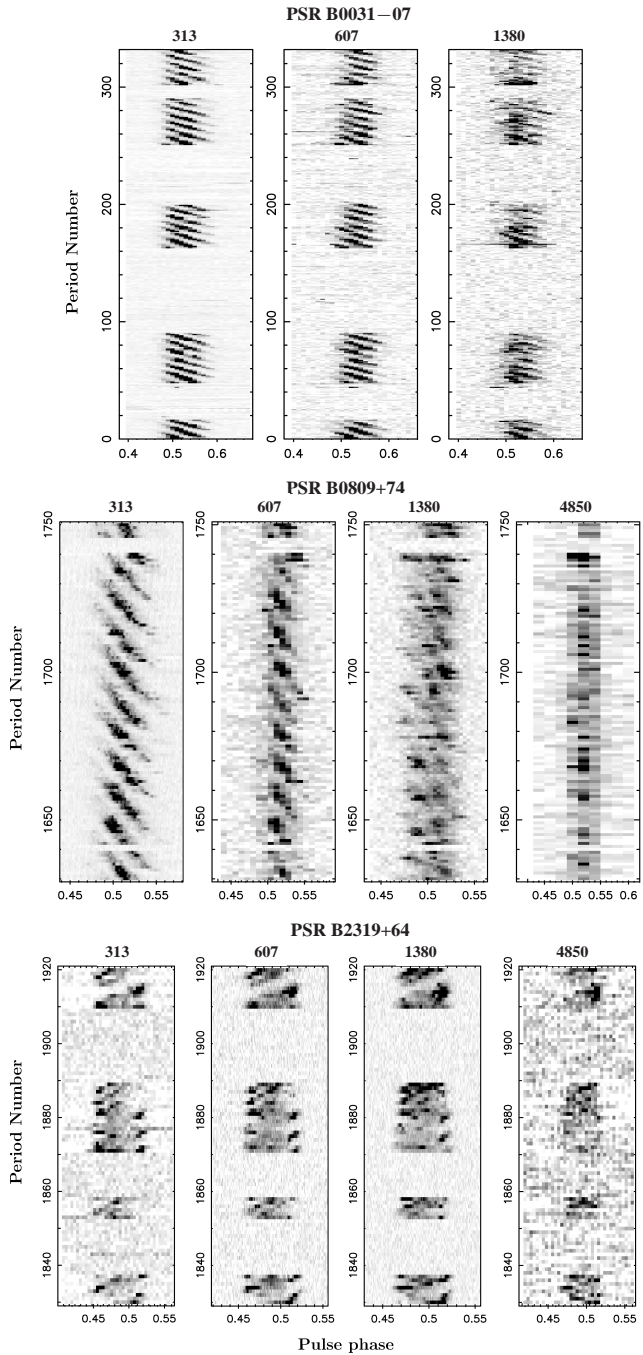}
\caption{Single pulse sequence, with gray scaled intensity 
as a function of pulse number and pulse phase, for a subset of 
data at all frequencies for PSRs B0031$-$07, 
B0809+74 and B2319+60. The sequences were observed 
simultaneously at all frequencies. Concurrent occurrence 
of null pulses is clearly evident for each pulsar.}
\normalfont
\label{spseq}
\end{center}
\end{figure}
 
The WSRT \citep{bh74} was also used in a tied array 
configuration of fourteen 25-m antennas covering 1380 MHz band with a 
160-MHz bandwidth. The short spacing configuration 
of WSRT was employed to keep the ionospheric effects 
minimum. The 160-MHz radio frequency band,  received 
with the Multi-Frequency Front-End (MFFE) at each 
antenna, were down-converted to an intermediate 
frequency (IF) and divided into eight 20-MHz subbands 
before digitization and addition in a tied array added 
module (TAAM). The data were acquired using an 
upgraded WSRT Pulsar Machine II (PuMa-II), which 
is a flexible cluster of 62 nodes capable of 
processing instantaneous observing bandwidth 
of 160 MHz \citep{kss08}. The data were recorded 
for each subband with a sampling time of 50 ns 
along with time-stamp derived from observatory's 
Hydrogen maser for about 8 hours. 
The shorter baselines at WSRT together with its 
high latitude and higher observing frequency 
allows  the tied array to remain phased for a 
longer time-scale than at lower frequencies 
used at the GMRT. Hence, a contiguous 
single pulse sequence was recorded at WSRT. 
The large data volume was reduced using off-line 
processing pipeline at WSRT to 8 dedispersed 
time-series with an effective sampling time of 1 
ms. 

The 100-m Effelsberg telescope was used at 4850 MHz 
with 500-MHz bandwidth in total intensity mode 
for two pulsars in the sample, PSRs B0809+74 and B2319+60. 
The detected signals from 6-cm dual horn secondary focus
receiver were acquired using PSRFFTS search
backend, configured to have 128 frequency channels
across 500-MHz bandpass, with a dump time of 64 $\mu$s 
and were recorded to a file along with 
a time-stamp derived from the observatory 
hydrogen maser.
 
Given the difference in the longitude of the 
observatories, the total overlap at all frequencies was 
smaller than the total duration of observations at 
each telescope. Part of the data during the overlap 
was affected by radio frequency interference (RFI)
at one or the other telescope and was not considered 
for the analysis described below. 

The data from all observatories were converted 
into a standard format required for 
SIGPROC\footnote{http://sigproc.sourceforge.net/} 
analysis package and dedispersed using the programs 
provided in the package. These were then folded to 
1000 bins across the period using the 
ephemeris of these pulsars using 
TEMPO\footnote{http://tempo.sourceforge.net/} 
package to obtain a single pulse sequence. 

First, the pulse sequence for the longest data file, 
typically consisting of 6000 pulses, was averaged  
for each frequency to obtain an integrated profile, 
which was used to form a noise-free template, 
after centering the pulse, for the pulsar 
at that frequency. The template at each frequency was 
used to estimate the number of samples to be removed 
from the beginning of each file for the observed frequencies 
so that the pulse is centered in a single period and 
time stamps for single pulses were corrected 
by these offsets. The single pulse sequences were then 
aligned by converting these time stamps to solar system 
barycentre (SSB) using TEMPO2$^2$. This conversion also takes into 
account the delay at lower frequencies due to dispersion 
in the inter-stellar medium (ISM). Then, the pulses 
corresponding to identical time stamps at SSB across 
all frequencies were extracted from the data. 
It should be noted that, the ISM also introduces further delay to 
the propagating signal due to multi-path propogation, which is responsible for prominent pulse broadening at lower radio frequencies 
for high DM pulsars. The NE2001 electron density model \cite[]{NE2001}  
predicts a delay of the order of $10^{-3}$ seconds 
for the highest DM pulsar in our sample. 
There is also a frequency dependent delay caused by the ionosphere which changes according 
to the time of the day. However, such delays are of the order of $10^{-7}$ seconds. Finally, there is a delay introduced by the Earth's atmosphere, 
which is of the order of 10 ns [See \cite{hem06} for details 
of these delays]. 
For the current observations, where the comparison is carried out 
on the time-scale of the periods of the pulsars ($\sim$1 second), 
such delays can be ignored as they are three to seven orders of 
magnitude smaller, respectively. 
As the observations were typically recorded 
in 2$-$3 data files at the GMRT (313 and 607 MHz),
the data recorded at WSRT and Effelsberg were split 
into similar number of files with observations 
duration equal to that at the GMRT.

The single pulse sequences were then visually examined to 
remove any single pulses with excessive RFI. 
The number of pulses available 
at all frequencies simultaneously 
after eliminating pulses affected by RFI are indicated 
in Table \ref{tabobs} for each pulsar. 
Nulling fractions (NFs) with their errors were obtained from 
the on-pulse energy and off-pulse energy sequences using 
a procedure similar to that described in \cite{gjk12}
for PSR B2319+60 and B0809+74. This method underestimates 
the NF for weaker pulsars with prominent sub-pulse drift 
as in the case of PSR B0031$-$07. Hence a method based on 
extreme values, described in \cite{viv95} was used 
for this pulsar. 

Sections of single pulse sequences with overlap on all 
frequencies and high S/N were 
extracted from the data as explained above. 
Null and burst pulses were identified using the 
threshold obtained from a histogram of on-pulse 
energy derived from on-pulse energy sequences,  
e.g. see Figure 1 in \cite{gjk12}. 
Any incorrectly identified null or burst pulse 
was relabeled after a careful visual examination of these 
plots to obtain a one-bit sequence with ones representing 
the bursts and zeros representing the nulls. Lengths 
of nulls (defined as a sequence of zeros bounded on both 
sides by bursts) and bursts were obtained for each frequency 
from the one-bit sequence and correlations between one-bit 
sequences across frequency were examined as explained 
in the next section. 

\section{Results} 
\label{res}
Single pulse sequences aligned in time across all 
frequencies for a section of data, observed 
simultaneously, are plotted for the three 
pulsars in Figure \ref{spseq}. All three pulsars 
show clear bursts interspersed with nulls. While 
PSRs B0031$-$07 and B2319+60 show frequent long nulls, 
PSR B0809+74 shows nulls up to 8 pulses separated 
by long sequence of burst pulses. Prominent drifting 
subpulses are seen in PSRs B0031$-$07 and B0809+74. 
The profiles of these pulsars (not shown) are known to 
evolve  with frequency in a manner consistent with 
a tangential traverse of line-of-sight to the emission 
beam. For PSR B2319+60, subpulse drift has been reported 
at the edges of its profile \citep{wf81}. 
The multi-component nature of this pulsar's 
profile implies a more central traverse of line-of-sight 
to the emission beam encompassing both the core and 
conal emission \cite[]{ran86,gl98}. 
\begin{figure}[h!]
\begin{center}
\includegraphics[angle=0,scale=.41]{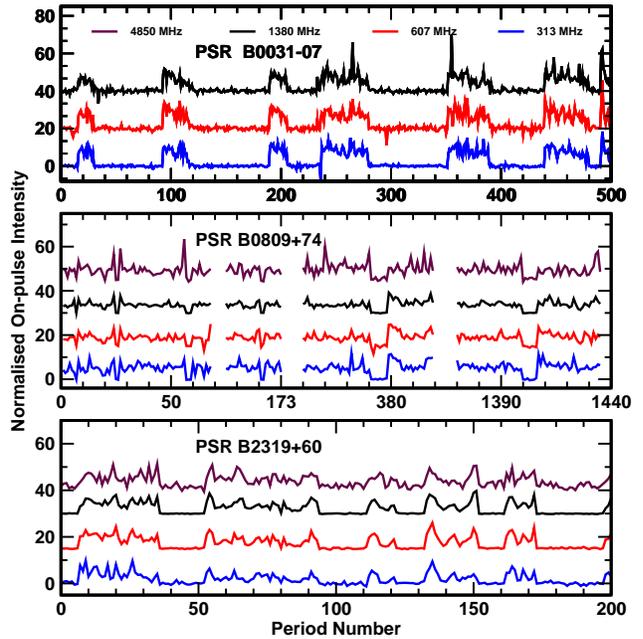}
\caption{On-pulse energy sequences as a function of pulse number for 
a subset of data for all frequencies for PSRs B0031$-$07, 
B0809+74 and B2319+60. For B0809+74, four sections 
of single pulses exclusively around the null phases are shown.} 
\label{figope}
\end{center}
\end{figure}

The pulsed emission switches simultaneously from burst state to null state 
(and vice versa) at all frequencies. This is true for both 
long nulls as well as short nulls (e.g. the null near 
pulse numbers 90 to 160 for PSR B0031$-$07 and 1638 to 
1645 for PSR B0809+74 in Figure \ref{spseq}). 
A visual inspection of the entire data broadly confirms this behavior. 
The pulse energy in the bins, where pulsed emission is 
present is shown in Figure \ref{figope} 
for the three pulsars, suggesting similar correlated 
behavior across the frequencies. Thus by visually inspecting the data, 
pulse nulling appears to be broadband over at least a decade of 
frequency for these pulsars. 

The NF, the distributions of null and burst lengths and 
the correlation between the nulling pattern represented 
by the one-bit sequence were compared across 
frequencies to quantify this correlated behavior. 
The NFs estimated for the three pulsars were found to 
be consistent within errors across the observed 
frequencies (Table \ref{tabnf}). The S/N was not sufficient to 
obtain a good estimate at 4850 MHz for PSR B0809+74 and 
B2319+60, so only a lower limit could be estimated in these 
cases. The combined NFs from all frequencies were estimated 
to be 42.8\% $\pm$ 0.6\%, 1.4\% $\pm$ 0.2\% and 30.5\% $\pm$ 0.6\% for PSRs B0031-07, 
B0809+74 and B2319+60, respectively. The expected error 
on the above mentioned NF was calculated as 
1.9$\times\sqrt{NF(1-NF)/P}$, assuming binomial distribution of null pulses. 
It should be noted that, the NF is measured as a ratio of total number 
of null pulses among $P$ observed pulses 
combining all frequency data for each pulsar. 
These estimates are reported here 
to highlight the consistency with the previously reported NF values 
for these pulsars \citep{viv95,la83,rit76}. 

The distributions of null and burst lengths, derived from the 
one-bit sequences, look similar for all frequencies 
for each pulsar indicating a very similar nulling 
pattern. These distributions were compared using a 
Kolmogorov-Smirnov test, which rejected, at a very 
high significance ($\geq$99.7\%), the hypothesis 
that these distributions for any pair of 
frequencies are different. 

Finally, the one-bit sequences for a pair of observing 
frequencies across all observations of each pulsar 
were compared using a contingency table analysis 
\citep{pftv86}. As these sequences represent 
two distinct states of a pulsar, the correlation between 
these states for a pair of frequency can be arranged as a 
2 $\times$ 2 contingency table. A $\phi$ test (Cramer-V) and 
uncertainty test based on entropy calculations can then be 
used to assess the strength and significance of any correlation. 
Both these tests result in a value between 0 and 1. A value 
close to 1 indicates a strong correlation. Cramer's V for a 
2 $\times$ 2 contingency table, as in our case, is just a measure 
of reduced chi-square. Hence, a value close to 1 indicates a 
very high significance of correlation. Likewise, a value 
close to unity for the uncertainty coefficient 
calculated from entropy arguments indicates a very 
high probability of observing null (burst) pulses at both 
frequencies. See \cite{pftv86} for details about these tests. The results 
of these tests are presented in Table \ref{tabcont}. Both Cramer-V 
and the uncertainty coefficients have values very close to 1 for 
PSRs B0031$-$07 and B0809+74 indicating a significant high 
association across all pairs of frequencies. This is also the case 
for PSR B2319+60, although the strength is marginally smaller 
for association between pairs involving 4850 MHz. 

\begin{table*}[h!]
\begin{center}
\caption{Estimated NFs of PSRs B0031$-$07,  B0809+74 and B2319+60. 
Columns give pulsar name, number of pulses used in the analysis and 
NF at each of the indicated observing frequency}
\label{tabnf}
\begin{tabular}{l|c|cccc}
\tableline\tableline
Pulsar         & Number of & \multicolumn{4}{c}{Frequency of Observations (MHz) } \\
               &  pulses& 303     & 607     & 1380  & 4850 \\
\tableline
PSR B0031$-$07 &  10441 & 43(2)   & 44(2)   & 43(2) & -     \\
PSR B0809+74   &  10003 & 1.4(3)  & 1.6(4)  & 1.2(2)& $>$ 1   \\
PSR B2319+60   &   5126 & 35(5)   & 33(3)   & 31(2) & $>$ 30  \\
\tableline
\end{tabular}
\end{center}
\end{table*}

\begin{table*}[h!]
\begin{center}
\caption{Estimate of correlation significance and strength for 
one-bit sequences between a pair of frequencies for  PSRs B0031-07,  
B0809+74 and B2319+60. The first row in each column for a given 
frequency gives the Cramer$-$V indicating the significance of correlation 
of one bit sequences associated with it and the frequency in the column. 
The second row in each column for a given frequency gives the 
corresponding uncertainty coefficient derived from entropy 
arguments (See \cite{pftv86})}
\label{tabcont}
\begin{tabular}{l|cc|ccc|ccc}
\tableline\tableline
     &\multicolumn{2}{c}{PSR B0031-07} & \multicolumn{3}{c}{PSR B0809+74} & \multicolumn{3}{c}{PSR B2319+60}\\
\tableline\tableline
Frequency of Observations (MHz) & 607  & 1380 & 607  & 1380 & 4850 &  607  & 1380 & 4850 \\
\tableline
313 & 0.99 & 0.99 & 0.96 & 0.97 & 0.96 & 0.98 & 0.98 & 0.94 \\
    & 0.96 & 0.95 & 0.92 & 0.92 & 0.91 & 0.91 & 0.93 & 0.81 \\
607 & $-$  & 0.99 & $-$  & 0.96 & 0.96 & $-$ & 0.98 & 0.94 \\
    & $-$  & 0.94 & $-$  & 0.91 & 0.90 & $-$ & 0.94  & 0.83 \\ 
1380 &     &  $-$ &      & $-$  & 0.97 &     & $-$   & 0.95  \\
    &      &  $-$ &      & $-$  & 0.92 &     & $-$   & 0.85 \\
\tableline\tableline
\end{tabular}
\end{center}
\end{table*}

For a small number of pulses, the above association does not 
hold. All such pulses were carefully examined for the three pulsars. 
About 44 pulses do not show simultaneous null (burst) 
state for PSR B0031$-$07. Among these, 25 occurred either at the start 
or at the end of a burst. There were 27 and 158 pulses observed in 
PSRs B0809+74 and B2319+60 respectively, where the state of 
pulse was not identical across the frequencies. Similar to 
PSR B0031$-$07, 20 and 82 of these occurred at the start/end of 
a burst for PSRs B0809+74 and B2319+60, respectively. Thus, while 
the nulling patterns for the three pulsars is largely broadband, 
deviations from this behavior is seen in about one to three percent 
of pulses, about half of which occur at the transition from 
null to burst (or vice versa). The fraction of such pulses 
is marginally higher for PSR B2319+60 than that 
for PSRs B0031$-$07 and B0809+74. We did not find any 
correlation across frequencies in the number of these 
non-concurrent pulses as they appear to be similar 
for different pairs of frequencies for all three pulsars. 

\section{Discussion}
\label{conc}
We find that nulling is broadband for PSRs B0031$-$07, B0809+74 and 
B2319+60 with their NFs consistent across the frequencies. 
The distributions of null and burst pulses were found to be similar and 
the null-burst sequences show significantly high association from 
313 to 4850 MHz. Deviation from this behavior 
was seen for less than 3\% of the pulses, most of which 
occur at the transition state (from null to burst or vice versa). 
For PSR B0809+74, we found highly concurrent broadband behavior for 
both long and short nulls, contrary to the reported behavior by 
\cite{bkk+81} and \cite{dls+84}. These studies were based on 
small number of pulses. It should be noted that our results 
are also different from those for  
PSR B1133+16, where 50\% of null pulses were reported to be 
non-concurrent \cite[]{bgk+07}.
It is likely that B1133+16 exhibits unique 
frequency dependent nulling behavior while the pulsars studied 
here exhibit frequency independent quenching of pulsed emission. 
Another explanation could be the frequency dependent pulse-to-pulse 
modulation in PSR B1133+16. 
\cite{bsw80} suggested that pulse energy modulation decreases 
upto a critical frequency ($\sim$ 1 GHz) and then increases. 
This dependence of the pulse energy modulation 
may result in non-concurrent behavior in pulsars such as PSR B1133+16.  
\cite{kkg+03} also speculated a loss in coherence across wide range 
of frequencies for this pulsar.

Among the two different group of theories, geometric models 
invoking empty line-of-sight ({\itshape pseudo-nulls}) are unlikely to apply for these pulsars 
as the emission is expected to be seen at some frequency for higher number of instances 
due to the frequency dependent separation between sub-beams. 
For PSR B1133+16 also the emission seen at higher frequencies, 
during the nulls at lower frequencies, was shown to be distinct from the normal emission,  
probably originating at higher heights \cite[]{bgk+07}. 
Moreover, as the sub-beams are arranged in a uniform 
pattern, unexcited emission regions imply some sort 
of periodicity, which is not seen in our data for these
pulsars. Changes in emission geometry as in movement of emission
region within the beam \cite[]{smk05} also implies that the emission
is expected to be seen at some frequency or other.  While this is not
ruled out, it is unlikely to be a general explanation for the nulling
phenomenon considering that we do not see this emission for both conal
pulsars such as PSRs B0031$-$07 and B0809+74 as well as for PSR B2319+60
which has a more central cut of the line of sight.

Intrinsic models invoking extinction of the sparking region or coherence conditions 
are consistent with our results. Such models propose the entire 
magnetosphere undergoing rapid changes to cause cessation of 
radio emission. \cite{klo+06} have shown cessation of emission 
to be associated with disruption of entire particle flow while 
\cite{lhk+10} have shown radio emission changes (mode-changing) 
occurring on the global magnetospheric scale.
Moreover, recent simultaneous observations of a mode-changing pulsar, 
PSR B0943+10, at radio and X-ray band also highlighted 
such global magnetospheric state switching \cite[]{hhk+13}. 
If the absence of emission during the null state is related 
to such intrinsic phenomena, the non-concurrent pulses 
may represent relaxation or local changes.  
The magnetospheric state switching may not be sudden 
and it is likely to posses finite relaxation time to 
switch between different states. For example, PSRs J1752+2359 
and J1738$-$2330 show gradual decay and slow rise in the pulse energy 
before and after the null state respectively \cite[]{gjw13}. 
This kind of gradual changes could manifest itself 
to give rise to differences in the emission state at different 
frequencies near the transition instances. 
Moreover, \cite{bgk+07} also reported lower frequencies to have longer nulls compared to nulls 
at higher frequencies in PSR B1133+16, pointing towards similar deviations at the transition instances. 
The distinct emission seen at higher frequencies for PSR B1133+16 can also be 
considered as a unique mode-changing phenomena in which one magnetospheric state 
does not produce detectable emission at lower frequencies. 
Thus, these results indicate that nulling seems to be another 
form of mode-change phenomena which invokes changes on the global magnetospheric scale, 
further supporting claims by \cite{klo+06} and \cite{lhk+10}. 

\acknowledgments
We thank the staff of the GMRT who made these observations possible. The
GMRT is operated by the National Centre for Radio Astrophysics of the Tata
Institute of Fundamental Research (NCRA-TIFR). We would also like to thank 
staff of the WSRT and the Effelsberg radio telescope to make these 
simultaneous observations possible. VG acknowledge the West Light Foundation of the
Chinese Academy of Sciences project XBBS-2014-21. We would also 
like to thank anonymous referee for the helpful comments.  

\bibliographystyle{apj}

\end{document}